\let\includefigures=\iftrue
\newfam\black

\input harvmac
\input rotate
\input rlepsf
\input xyv2
\noblackbox
\includefigures
\message{If you do not have epsf.tex (to include figures),}
\message{change the option at the top of the tex file.}
\def\figin{\epsfcheck\figin}\def\figins{\epsfcheck\figins}
\def\epsfcheck{\ifx\epsfbox\UnDeFiNeD
\message{(NO epsf.tex, FIGURES WILL BE IGNORED)}
\gdef\figin##1{\vskip2in}\gdef\figins##1{\hskip.5in}
\else\message{(FIGURES WILL BE INCLUDED)}%
\gdef\figin##1{##1}\gdef\figins##1{##1}\fi}
\def\DefWarn#1{}

\def\figinsert{\goodbreak\midinsert}
\def\ifig#1#2#3{\DefWarn#1\xdef#1{fig.~\the\figno}
\writedef{#1\leftbracket fig.\noexpand~\the\figno}%
\figinsert\figin{\centerline{#3}}\medskip\centerline{\vbox{\baselineskip12pt
\advance\hsize by -1truein\noindent\footnotefont{\bf
Fig.~\the\figno:} #2}}
\bigskip\endinsert\global\advance\figno by1}
\else
\def\ifig#1#2#3{\xdef#1{fig.~\the\figno}
\writedef{#1\leftbracket fig.\noexpand~\the\figno}%
\global\advance\figno by1} \fi
%
%
\lref\DijkgraafFC{ R.~Dijkgraaf and C.~Vafa, ``Matrix models,
topological strings, and supersymmetric gauge theories,'' Nucl.\
Phys.\ B {\bf 644}, 3 (2002) [arXiv:hep-th/0206255].
}
\lref\DijkgraafVW{ R.~Dijkgraaf and C.~Vafa, ``On geometry and
matrix models,'' Nucl.\ Phys.\ B {\bf 644}, 21 (2002)
[arXiv:hep-th/0207106].
}
\lref\DijkgraafDH{ R.~Dijkgraaf and C.~Vafa, ``A perturbative
window into non-perturbative physics,'' arXiv:hep-th/0208048.
}
\lref\CachazoJY{ F.~Cachazo, K.~A.~Intriligator and C.~Vafa, ``A
large N duality via a geometric transition,'' Nucl.\ Phys.\ B {\bf
603}, 3 (2001) [arXiv:hep-th/0103067].
}
\lref\CachazoPR{ F.~Cachazo and C.~Vafa, ``N = 1 and N = 2
geometry from fluxes,'' arXiv:hep-th/0206017.
}
\lref\CachazoRY{ F.~Cachazo, M.~R.~Douglas, N.~Seiberg and
E.~Witten, ``Chiral rings and anomalies in supersymmetric gauge
theory,'' JHEP {\bf 0212}, 071 (2002) [arXiv:hep-th/0211170].
}
\lref\CachazoZK{ F.~Cachazo, N.~Seiberg and E.~Witten, ``Phases of
N = 1 supersymmetric gauge theories and matrices,'' JHEP {\bf
0302}, 042 (2003) [arXiv:hep-th/0301006].
}
\lref\CachazoYC{ F.~Cachazo, N.~Seiberg and E.~Witten, ``Chiral
Rings and Phases of Supersymmetric Gauge Theories,'' JHEP {\bf
0304}, 018 (2003) [arXiv:hep-th/0303207].
}
\lref\FerrariJP{ F.~Ferrari, ``On exact superpotentials in
confining vacua,'' Nucl.\ Phys.\ B {\bf 648}, 161 (2003)
[arXiv:hep-th/0210135].
}
\lref\FerrariKQ{ F.~Ferrari, ``Quantum parameter space and double
scaling limits in N = 1 super  Yang-Mills theory,'' Phys.\ Rev.\ D
{\bf 67}, 085013 (2003) [arXiv:hep-th/0211069].
}
\lref\FerrariYR{ F.~Ferrari, ``Quantum parameter space in super
Yang-Mills. II,'' Phys.\ Lett.\ B {\bf 557}, 290 (2003)
[arXiv:hep-th/0301157].
}
\lref\DijkgraafXD{ R.~Dijkgraaf, M.~T.~Grisaru, C.~S.~Lam, C.~Vafa
and D.~Zanon, ``Perturbative computation of glueball
superpotentials,'' arXiv:hep-th/0211017.
}
\lref\VenezianoAH{ G.~Veneziano and S.~Yankielowicz, ``An
Effective Lagrangian For The Pure N=1 Supersymmetric Yang-Mills
Theory,'' Phys.\ Lett.\ B {\bf 113}, 231 (1982).
}
\lref\SeibergRS{ N.~Seiberg and E.~Witten, ``Electric - magnetic
duality, monopole condensation, and confinement in N=2
supersymmetric Yang-Mills theory,'' Nucl.\ Phys.\ B {\bf 426}, 19
(1994) [Erratum-ibid.\ B {\bf 430}, 485 (1994)]
[arXiv:hep-th/9407087].
}
\lref\SeibergAJ{ N.~Seiberg and E.~Witten, ``Monopoles, duality
and chiral symmetry breaking in N=2 supersymmetric QCD,'' Nucl.\
Phys.\ B {\bf 431}, 484 (1994) [arXiv:hep-th/9408099].
}
\lref\ArgyresXH{ P.~C.~Argyres and A.~E.~Faraggi, ``The vacuum
structure and spectrum of N=2 supersymmetric SU(n) gauge theory,''
Phys.\ Rev.\ Lett.\  {\bf 74}, 3931 (1995) [arXiv:hep-th/9411057].
}
\lref\KlemmQS{ A.~Klemm, W.~Lerche, S.~Yankielowicz and
S.~Theisen, ``Simple singularities and N=2 supersymmetric
Yang-Mills theory,'' Phys.\ Lett.\ B {\bf 344}, 169 (1995)
[arXiv:hep-th/9411048].
}
\lref\HananyNA{ A.~Hanany and Y.~Oz, ``On the quantum moduli space
of vacua of N=2 supersymmetric SU(N(c)) gauge theories,'' Nucl.\
Phys.\ B {\bf 452}, 283 (1995) [arXiv:hep-th/9505075].
}
\lref\DouglasNW{ M.~R.~Douglas and S.~H.~Shenker, ``Dynamics of
SU(N) supersymmetric gauge theory,'' Nucl.\ Phys.\ B {\bf 447},
271 (1995) [arXiv:hep-th/9503163].
}
\lref\ArgyresJJ{ P.~C.~Argyres and M.~R.~Douglas, ``New phenomena
in SU(3) supersymmetric gauge theory,'' Nucl.\ Phys.\ B {\bf 448},
93 (1995) [arXiv:hep-th/9505062].
}
\lref\ArgyresXN{ P.~C.~Argyres, M.~Ronen Plesser, N.~Seiberg and
E.~Witten, ``New N=2 Superconformal Field Theories in Four
Dimensions,'' Nucl.\ Phys.\ B {\bf 461}, 71 (1996)
[arXiv:hep-th/9511154].
}
\lref\EguchiVU{ T.~Eguchi, K.~Hori, K.~Ito and S.~K.~Yang, ``Study
of $N=2$ Superconformal Field Theories in $4$ Dimensions,'' Nucl.\
Phys.\ B {\bf 471}, 430 (1996) [arXiv:hep-th/9603002].
}
\lref\TerashimaPN{ S.~Terashima and S.~K.~Yang,``Confining phase
of N = 1 supersymmetric gauge theories and N = 2  massless
solitons,'' Phys.\ Lett.\ B {\bf 391}, 107 (1997)
[arXiv:hep-th/9607151].
}
\lref\TerashimaZV{ S.~Terashima and S.~K.~Yang, ``ADE confining
phase superpotentials,'' Nucl.\ Phys.\ B {\bf 519}, 453 (1998)
[arXiv:hep-th/9706076].
}
\lref\EguchiWV{ T.~Eguchi and Y.~Sugawara, ``Branches of N = 1
vacua and Argyres-Douglas points,'' JHEP {\bf 0305}, 063 (2003)
[arXiv:hep-th/0305050].
}
\lref\BertoldiAB{ G.~Bertoldi, ``Matrix models, Argyres-Douglas
singularities and double scaling limits,'' JHEP {\bf 0306}, 027
(2003) [arXiv:hep-th/0305058].
}
\lref\GorskyEJ{ A.~Gorsky, A.~I.~Vainshtein and A.~Yung,
Nucl.\ Phys.\ B {\bf 584}, 197 (2000) [arXiv:hep-th/0004087].
}
\lref\deBoerAP{ J.~de Boer and Y.~Oz,
Nucl.\ Phys.\ B {\bf 511}, 155 (1998) [arXiv:hep-th/9708044].
}


\def\IL{\relax{\rm I\kern-.18em L}}
\def\IH{\relax{\rm I\kern-.18em H}}
\def\IR{\relax{\rm I\kern-.18em R}}
\def\IC{\relax\hbox{$\inbar\kern-.3em{\rm C}$}}
\def\IZ{\relax\ifmmode\mathchoice
{\hbox{\cmss Z\kern-.4em Z}}{\hbox{\cmss Z\kern-.4em Z}}
{\lower.9pt\hbox{\cmsss Z\kern-.4em Z}} {\lower1.2pt\hbox{\cmsss
Z\kern-.4em Z}}\else{\cmss Z\kern-.4em Z}\fi}

\def\CN {{\cal N}}

\def\CF {{\cal F}}

\def\CO {{\cal O}}


\def\CN {{\cal N}}

\def\CO {{\cal O}}

\def\Tr{{\rm Tr}}

\def\lfm#1{\medskip\noindent\item{#1}}
\font\manual=manfnt \def\dbend{\lower3.5pt\hbox{\manual\char127}}

\def\IZ{\relax\ifmmode\mathchoice
{\hbox{\cmss Z\kern-.4em Z}}{\hbox{\cmss Z\kern-.4em Z}}
{\lower.9pt\hbox{\cmsss Z\kern-.4em Z}} {\lower1.2pt\hbox{\cmsss
Z\kern-.4em Z}}\else{\cmss Z\kern-.4em Z}\fi}

\def\rt2{\sqrt{2}}
\def\irt2{{1\over\sqrt{2}}}

\def\hat{\widehat}
\writedefs
%
%
%
%
\newbox\tmpbox\setbox\tmpbox\hbox{\abstractfont }
\Title{\vbox{\baselineskip12pt
\hbox{hep-th/0308001}\hbox{PUPT-2092}
}}
{\vbox{\centerline{Singularities of $\CN=1$ Supersymmetric Gauge
Theory}\smallskip\centerline{and Matrix Models}}}
\smallskip
\centerline{David Shih}
\smallskip
\bigskip
\centerline{Department of Physics, Princeton University, Princeton, NJ 08544}
\medskip
\bigskip
\vskip 1cm
\noindent
In $\CN=1$ supersymmetric $U(N)$ gauge theory with adjoint matter $\Phi$
and polynomial tree-level superpotential $W(\Phi)$, the massless fluctuations
about each quantum vacuum are generically described by $U(1)^n$ gauge theory
for some $n$. However, by tuning the parameters of $W(\Phi)$ to
non-generic values, we can reach singular vacua where additional fields
become massless. Using both the matrix model prescription and the strong-coupling approach,
we study in detail three examples of such singularities: the singularities of the $n=1$
branch, intersections of $n=1$ and $n=2$ branches, and a class
of $\CN=1$ Argyres-Douglas points. In all three examples, we find
that the matrix model description of the low-energy physics
breaks down in some way at the singularity.

\Date{August, 2003}

\newsec{Introduction}

Following the work of Dijkgraaf and Vafa
\refs{\DijkgraafFC\DijkgraafVW-\DijkgraafDH}, much recent progress
has been made in the study of a wide class of supersymmetric gauge
theories and their connections to bosonic matrix models. Probably
the simplest and most well-studied example in this context (see
e.g.~\refs{\CachazoRY\CachazoZK\CachazoYC\FerrariJP\FerrariKQ\FerrariYR-
\DijkgraafXD}) has been $\CN=1$ supersymmetric $U(N)$ gauge
theory, with matter $\Phi$ in the adjoint representation and
polynomial tree-level superpotential
\eqn\Wtreedef{
    W(\Phi)=\sum_{r=0}^{k}{g_{r}\over r+1}\,\Tr\,\Phi^{r+1}.
}
The associated matrix model in this case is a zero-dimensional
model of an $\hat{N}\times \hat{N}$ complex matrix $M$, with
potential $W(M)$. The idea of
\refs{\DijkgraafFC\DijkgraafVW-\DijkgraafDH} was that the
$\hat{N}\rightarrow \infty$ limit of the matrix model could be
used to obtain certain exact, non-perturbative gauge theory
quantities, in a low-energy description of the gauge theory that
we will call the ``glueball description." It will be useful to
briefly review the semiclassical arguments that lead to the
glueball description. For generic tree-level superpotential,
$\Phi$ acquires a mass and an expectation value, breaking $U(N)$
down to $G=\Pi_{i=1}^{n}U(N_i)\cong U(1)^n\times\Pi_{i=1}^n
SU(N_i)$ for some $n\le k$. Integrating out $\Phi$ and the massive
$W$-bosons leads to an effective theory of pure super-Yang-Mills
with gauge group $G$. At lower energies, the $SU(N_i)$ factors are
believed to confine and become massive, leaving behind only a
$U(1)^n$ gauge theory in the far infrared. However, it is also
believed that the glueball fields $S_i$ of the $SU(N_i)$ factors
behave as elementary fields in the IR \VenezianoAH. These $n$
massive glueball fields together with the $U(1)^n$ photons
comprise the glueball description of the low-energy gauge theory.

The quantities computed by the associated matrix model are the
effective superpotential $W_{eff}(S_i)$ for the $n$ glueball
fields and the matrix of low-energy $U(1)^n$ gauge couplings. The
effective glueball superpotential is the source of much
interesting physics. For instance, extremizing $W_{eff}$ leads to
the exact, non-perturbative, quantum vacua of the original gauge
theory. Also the quadratic part of $W_{eff}$ indicates whether the
$S_i$ are massive or not, if one also assumes regularity of the
K\"ahler potential. For generic tree-level superpotentials, we
expect the matrix model to provide a good description of the
low-energy physics, in the sense that it reproduces the exact
non-perturbative vacuum structure and gives the correct massless
spectrum (a number of $U(1)$ multiplets).

In this paper, we will be interested in how the generic picture
can be modified at strong-coupling singularities where additional
fields become massless. These singularities are reached by tuning
one or more parameters of $W(\Phi)$ to special values, and are
thus non-generic. In the following sections, we will study some
examples of such singularities. The examples are certainly worth
studying for their own sake, as the presence of extra massless
fields can lead to novel low-energy physics such as massless
monopoles and interacting SCFTs. In addition, the study of
singularities will teach us about the limitations of the gauge
theory/matrix model correspondence, since we expect the glueball
description to break down at singularities. The reason for this is
that the extra fields that become massless at a singularity can be
thought of as fields (i.e.~monopoles and the Casimirs
$\Tr\,\Phi^k$) in the ``strong-coupling description" of the
low-energy theory \refs{\CachazoPR, \CachazoJY} based on the exact
Seiberg-Witten solution of the underlying $\CN=2$ gauge theory
\refs{\SeibergRS\SeibergAJ\ArgyresXH\KlemmQS-\HananyNA}.\foot{We
will assume that the strong-coupling description is applicable
even at singularities. This is possible for the following reason.
If we write $W'(x)=g\,\Pi_{i=1}^{k}(x-a_i)$, then the
strong-coupling approach is always valid for $g$ sufficiently
small. Meanwhile, singular behavior is generally controlled by the
values of the $a_i$. Therefore, we can tune the $a_i$ to reach a
singularity, while keeping $g$ sufficiently small to ensure that
the strong-coupling description remains valid.} Thus they are not
included in the glueball description. Rather, they have been
integrated out, and integrating out massless fields usually leads
to a breakdown in an effective theory. Through our study of
examples, we will see various ways in which the breakdown of the
glueball description is manifested. At the same time, it is
important to keep in mind that the extrema of the effective
glueball superpotential should still determine the vacua of the
theory even when the glueball description itself breaks down. This
is guaranteed by holomorphy.

The outline of the paper is as follows. We study the three
examples described below in sections 2--4, respectively. Appendix
A contains some calculations relevant to the study of the
$n=1$/$n=2$ singularity in section 2. Appendix B describes the
matrix model calculation of the effective glueball superpotential
relevant to the discussion of the Argyres-Douglas (AD) point in
section 3. The three examples we will study are:

 \lfm{1.} Singularities of the $n=1$ branch. We will study the
$n=1$ branch for tree-level superpotential of arbitrary degree,
generalizing the results of \refs{\FerrariKQ}. We will show that
the only singularities are those at which $\Tr\,\Phi$ becomes
massless in the strong-coupling description. At such points, the
mass of the glueball field $S$ also goes to zero, indicating a
breakdown of the glueball description. We will also discuss other
pathologies of the glueball description where the effective
glueball superpotential becomes non-analytic. This is taken to be
evidence that the glueball description has, in some sense, a
finite radius of convergence. We argue that at such pathological
points, the K\"ahler potential for $S$ must also degenerate.

\lfm{2.} Intersections of $n=1$ and $n=2$ branches for cubic
tree-level superpotential. These singularities were first
considered in \refs{\CachazoZK, \FerrariKQ}. They occur when a
vacuum with unbroken gauge group $U(N_1)\times U(N_2)$ meets a
vacuum with unbroken gauge group $U(N)$. The singularity arises
from an additional monopole becoming massless in the
strong-coupling description. We will solve the glueball equations
of motion exactly at the $n=1$/$n=2$ singularity, which will lead
to general formulas (functions of $N_1$ and $N_2$) for the
parameters of $W(\Phi)$ and the vevs of chiral operators at these
singularities. Previously, the $n=1$/$n=2$ singularities were
studied only for examples of $U(3)$--$U(6)$ in \CachazoZK. Our
general analysis significantly extends these results. In the
process, we confirm that extremizing the effective glueball
superpotential leads to the correct vacua even at singularities
where the glueball description breaks down. We also show that the
breakdown of the glueball description at the $n=1$/$n=2$
singularity occurs through a logarithmic divergence in the
low-energy gauge coupling. This divergence is to be expected from
the strong-coupling description, where a charged monopole field
becomes massless.

\lfm{3.} $\CN=1$ Argyres-Douglas points with
$W(\Phi)=\Tr\,\Phi^{n+1}$. Here, the singularity results from a
number of mutually non-local monopoles becoming massless. We will
focus on the case $n=n_{min}=\left[{N+1\over2}\right]$; $\CN=1$ AD
points with $n>n_{min}$ have been studied recently in
\refs{\EguchiWV, \BertoldiAB}. (The cases with $n<n_{min}$ are
uninteresting, since they does not lead to $\CN=1$ AD points.)
There has been some question as to whether the points with
$n=n_{min}$ are actually singular. We will compute the effective
superpotential in both the glueball and the strong-coupling
descriptions in an attempt to answer this question. We will
confirm that for $N$ even, the points with $n=n_{min}$ are
non-singular; while for $N$ odd we will find evidence that these
points lead to an interacting IR SCFT . The evidence takes the
form of a non-analytic strong-coupling superpotential.
Interestingly, the glueball superpotential remains analytic,
although a glueball becomes massless. Thus, as in the
singularities of the $n=1$ branch, a massless glueball again
signifies a breakdown in the glueball description due to the
presence of additional massless fields. However in this case, the
massless fields are mutually non-local monopoles.

\newsec{Singularities of the $n=1$ branch}

The first set of singularities we will study are those of the
$n=1$ branch. These singularities were previously discussed in
\refs{\FerrariKQ} for cubic tree-level superpotential. Our results
generalize the discussion to superpotentials of arbitrary degree,
and also hopefully clarify some points of confusion. We follow the
approach and conventions of \CachazoRY. Let the tree-level
superpotential be given by \Wtreedef. For a given $W(\Phi)$, there
are many quantum vacua, each associated with a choice of
semi-classical gauge group $\Pi_i U(N_i)$. The vacua corresponding
to unbroken gauge group $U(N)$ sweep out the $n=1$ branch as the
parameters of $W(\Phi)$ are varied. On the $n=1$ branch, there is
only one glueball field $S$ in the glueball description.
Similarly, in the strong coupling description there is only one
independent field $z_0={1\over N}\Tr\,\Phi$ remaining after the
$N-1$ condensed monopoles have been integrated out. The effective
superpotential for $z_0$ can be obtained exactly with the help of
Chebyshev polynomials \DouglasNW. The result is \CachazoRY:
\eqn\weffz{
    W_{eff}(z_0)=N\sum_{l=0}^{\left[{k+1\over2}\right]}{\Lambda^{2l}\over
    \left(l!\right)^2}W^{(2l)}(z_0).
}
The vev of $z_0$ is determined by solving the following polynomial
equation for $z_0$:
\eqn\weffzmin{
    W_{eff}'(\langle z_0\rangle)=N\sum_{l=0}^{\left[{k+1\over2}\right]}
    {\Lambda^{2l}\over \left(l!\right)^2}W^{(2l+1)}(\langle
    z_0\rangle)=0.
}
Since $W_{eff}(z_0)$ is polynomial in $z_0$, it is always finite
and regular near its extrema. Therefore, assuming regularity of
the K\"ahler potential (which is a valid assumption at least for
$W(\Phi)$ sufficiently small), the only possible singularities on
the $n=1$ branch must occur at points where $z_0$ becomes
massless. The condition for massless $z_0$ is simply
\eqn\weffzmassless{
    M_0\equiv W_{eff}''(\langle z_0\rangle)=N\sum_{l=0}^{\left[{k-1\over2}\right]}
            {\Lambda^{2l}\over \left(l!\right)^2}W^{(2l+2)}(\langle
            z_0\rangle)=0.
}
We can think of \weffzmassless\ as a restriction on the form of
the tree-level superpotential. The tree-level superpotential
originally has $k+1$ parameters, $(g_0,\dots,g_k)$. After using
\weffzmin\ to write $\langle z_0 \rangle$ in terms of the $g_i$,
\weffzmassless\ becomes one equation relating these parameters.
Therefore the massless $z_0$ singularities comprise a $k$
dimensional subspace of the $k+1$ dimensional parameter space of
tree-level superpotentials.

Now let us see how the massless $z_0$ singularities are realized
in the glueball description. Treating the glueball field as an
elementary degree of freedom in the IR depends on being able to
integrate out all the components of the adjoint field $\Phi$,
including $z_0$. Thus when $z_0$ is massless, the glueball
analysis should break down somehow. The effective glueball
superpotential can be obtained through a Legendre transform of
\weffz\ with respect to $2N\log\Lambda$ \CachazoRY. The result of
the Legendre transform is an effective superpotential for $S$,
$z_0$ and an auxiliary field $C$:
\eqn\weffnone{ W_{eff}(S,C,z_0)=2NS\log\left({\Lambda\over C}\right)+
N\sum_{l=0}^{\left[{k+1\over2}\right]}{C^{2l}\over(l!)^2}W^{(2l)}(z_0).
}
Integrating out $C$ and $S$ using their equations of motion
returns us to \weffz, while integrating out $C$ and $z_0$ gives
the effective glueball superpotential. The equations of motion for
$C$ and $z_0$ are:
\eqn\Candz{\eqalign{
    &\sum_{l=0}^{\left[{k+1\over2}\right]}{C^{2l}\over(l!)^2}W^{(2l+1)}(z_0)=0\cr
    &\sum_{l=1}^{\left[{k+1\over2}\right]}{l C^{2l}\over (l!)^2}W^{(2l)}(z_0)=S.\cr
}}
Solving for $C$ and $z_0$ and substituting back into \weffnone\
results in an effective superpotential for $S$, i.e.
\eqn\weffS{
W_{eff}(S)=W_{eff}(S,C(S),z_0(S))
}
with $C(S)$ and $z_0(S)$ determined implicitly via \Candz.

While it is not obvious that there exists a closed-form expression
for $W_{eff}(S)$, one can easily show that the first derivative of
$W_{eff}(S)$ is given by the very simple expression \eqn\weffSd{
W_{eff}'(S)=2N\log\left({\Lambda\over C(S)}\right). } Moreover,
using \Candz, one can show that the second derivative of $W_{eff}$
is \eqn\weffSdd{ W_{eff}''(S)={NA \over B^2-C^2 A^2} } with
\eqn\ABdef{\eqalign{
    A &=\sum_{l=0}^{\left[{k-1\over2}\right]}{C^{2l}\over (l!)^2}W^{(2l+2)}(z_0)\cr
    B &=\sum_{l=1}^{\left[{k+1\over2}\right]}{l C^{2l}\over (l!)^2} W^{(2l+1)}(z_0).\cr }
}
From \weffSd, we see that the extrema of the effective
superpotential occur when $C(S)=\Lambda$. Substituting this in
\ABdef\ and comparing with \weffzmassless, we see that $N\langle
A\rangle = M_0$. Thus at an extremum, \weffSdd\ becomes
\eqn\Smass{
    W_{eff}''(\langle S\rangle)= {N^2 M_0\over
    N^2\langle B\rangle^2-\Lambda^2 M_0^2}.
}
It is clear from \Smass\ that the glueball $S$ can become massless
only when the mass $M_0$ of $z_0$ also goes to zero. Examples of
such singularities for the case of cubic superpotential $k=2$ were
discussed in \refs{\FerrariKQ}. Our general analysis makes it
clear that there are no genuinely massless glueball points: since
the fundamental field $z_0$ also becomes massless, it is incorrect
to treat the composite glueball field as an elementary excitation
of the low-energy theory. Rather, we should think of the massless
glueball as a sign that the glueball description has broken down.

We should also consider the possibility that the denominator of
\Smass\ vanishes. This pathological behavior can occur for either
$z_0$ massive or massless, and it can result in an ``infinite
mass" for the glueball field. But the fact that it is not
correlated with the mass of $z_0$, together with the fact that the
strong-coupling description in terms of $z_0$ is comparatively
well-behaved, suggests that these pathological points should not
be thought of as additional singularities. Rather, we should
interpret them as evidence for the finite radius of convergence of
the glueball description. At such points it is reasonable to
suppose that the K\"ahler potential for $S$ also degenerates in
such a way that the entire pathology can be removed with a field
redefinition, so that the redefined glueball field is massive.
With this plausible assumption, the only true singularities of the
low-energy theory where additional fields become massless are the
massless $z_0$ points.

\newsec{The $n=1$/$n=2$ singularity}

\subsec{Solving the glueball equations of motion}

The second class of singularities we will study are the
intersections of the $n=1$ and $n=2$ branches. These singularities
occur at special values of the tree-level superpotential
parameters where a vacuum with unbroken gauge group $U(N_1)\times
U(N_2)$ meets a vacuum with unbroken gauge group $U(N)$. The basic
features of such singularities were discussed in \CachazoZK\ using
the strong-coupling approach. In this section, we will study the
intersection singularity in greater detail using the glueball
description. We will focus on studying the approach to the
singularity from the $n=2$ branch, because the corresponding
approach from the $n=1$ branch generally exhibits no singular
behavior. This can be seen in a variety of ways. For example, from
the previous section we know that the only singularities of the
$n=1$ branch are those associated with massless $z_0$.
Alternatively, the matrix model curve on the $n=1$ branch is
perfectly regular as we pass through the intersection with the
$n=2$ branch: it always has a single branch cut.

We will take the tree-level superpotential to be cubic to simplify
the calculations. The qualitative features of the analysis,
however, should be common to $n=1/n=2$ singularities for
superpotentials of arbitrary degree. As we approach the
$n=1$/$n=2$ singularity from the $n=2$ branch, both the matrix
model curve and the $\CN=2$ curve acquire an extra double root.
Using the matrix model curve, we can calculate the effective
glueball superpotential and the matrix of $U(1)$ gauge couplings
near the singularity. The matrix model curve takes the form
\eqn\ymarb{
    y_m^2=(x^2-m^2)^2+f_1 x+f_0=(x-r_1)(x-r_2)(x-r_3)(x-r_4).
}
Note that we have lost no
generality by taking the tree-level superpotential to have the
form $W'(x)=x^2-m^2$. The most general cubic superpotential can
always be recovered by a shift in $x$.

The first parametrization of the matrix model curve makes it clear
that $m$ is a parameter and $f_0$ and $f_1$ are fluctuating fields
related to the two glueball fields $S_1$ and $S_2$ by a field
redefinition. However, the second parametrization in terms of the
roots $r_i$ of $y_m$ will prove more convenient in the
calculations to follow. We should think of the four $r_i$ as
fields subject to the two constraints that the highest order terms
in \ymarb\ be given by $W'(x)^2$. We can always return to the
first interpretation involving $f_0$ and $f_1$ by matching the two
parameterizations of \ymarb.

To compute the effective glueball superpotential, we must evaluate
the period integrals of the matrix model curve. This results in a
complicated and not very illuminating combination of elliptic
integrals. Fortunately, however, the first derivatives with
respect to $f_j$ -- and hence the glueball equations of motion --
are relatively simple to evaluate. The calculation is briefly
outlined in appendix A; here we simply exhibit the final result
(dropping as usual irrelevant terms of $\CO({1\over \Lambda_0}))$:
\eqn\weffdereval{\eqalign{
    {\partial W_{eff}\over\partial f_0}&=
    {1\over\sqrt{(r_4-r_2)(r_3-r_1)}}\left(N F(\theta|R)-N_2F({\pi\over2}|R)
    \right)\cr
    {\partial W_{eff}\over\partial f_1}&=
    {1\over2}N\log\left({r_4+r_3-r_2-r_1\over 4\Lambda}\right)
    +{r_1\over\sqrt{(r_4-r_2)(r_3-r_1)}}\left(N F(\theta|R)-N_2 F({\pi\over2}|R)\right)\cr
    &\qquad +{(r_2-r_1)\over\sqrt{(r_4-r_2)(r_3-r_1)}}\left(N
    \Pi(n;\theta|R)-N_2\Pi(n;{\pi\over2}|R)\right)\cr
}}
where
\eqn\thetaRn{\eqalign{
    &\sin^2\theta = {r_4-r_2\over r_4-r_1},\quad R={(r_3-r_2)(r_4-r_1)\over
        (r_3-r_1)(r_4-r_2)},\quad n={r_3-r_2\over r_3-r_1}
}}
and $F$ and $\Pi$ are the standard elliptic integrals of the first
and third kinds, respectively. Their explicit definitions are
given in appendix A.\foot{We assume throughout this section that
the integer $b_2=0$. The effect of nonzero $b_2$ is easily
included by adding to \weffdereval\ the formula for the
derivatives of $S_2$ given in appendix A. In any case, a trivial
calculation shows that $b_2=0$ near an $n=1$/$n=2$ singularity.}

Eqn.\ \weffdereval\ is very interesting, as it represents a
general formula for the derivatives of the two-cut effective
superpotential, valid for every choice of $N_1$ and $N_2$. Let us
mention a couple of caveats, however. First, as mentioned in the
footnote, we have assumed $b_2=0$ for simplicity; the effects of $b_2\ne 0$
can be easily included. Secondly, we have made the
crucial ansatz that the $r_i$ lie on the real line. With this
ansatz, we can assume without loss of generality that
$r_1<r_2<r_3<r_4$, and we can choose the cuts of $y_m$ to be
$(r_1,r_2)$ and $(r_3,r_4)$. This is the starting point of the
calculations leading to \weffdereval\ described in appendix A.
Only with these assumptions about the $r_i$ are the elliptic
integrals in \weffdereval\ are unambiguously defined. One might
hope to analytically continue \weffdereval\ to arbitrary complex
$r_i$, but this procedure probably suffers from various
ambiguities. Perhaps with a little more effort these ambiguities
can be brought under control, but we postpone this for future
work. Fortunately, it is not relevant for the $n=1$/$n=2$
singularity.

The equations of motion for $f_0$ and $f_1$ are obtained by
setting the derivatives \weffdereval\ to zero. We can rewrite them
more simply as:
\eqn\eom{\eqalign{
    & N F(\theta|R)-N_2 F({\pi\over2}|R)=0\cr
    & N\log\left({r_4+r_3-r_2-r_1\over4\Lambda}\right)=
        -{2(r_2-r_1)\over\sqrt{(r_4-r_2)(r_3-r_1)}}\left(N\Pi(n;\theta|R)-N_2\Pi(n;{\pi\over2}|R)\right).\cr
}}
Notice that the equations of motion are most naturally expressed
in terms of the roots $r_i$ of the matrix model curve. As
discussed above, these roots are subject to two constraints:
\eqn\constraints{\eqalign{
    & r_1+r_2+r_3+r_4=0\cr
    &(r_2-r_3)^2+(r_1-r_4)^2-2(r_2+r_3)(r_1+r_4)=8m^2.\cr
}}
These constraints ensure that the highest-order terms in $y_m^2$
are given by $W'(x)^2$. Combining \eom\ with \constraints\ allows
us to solve for the $r_i$ in terms of the parameters $N_1$, $N_2$
and $m$. Matching the two parameterizations of the matrix model
curve \ymarb\ yields the expectation values of the fields $f_0$
and $f_1$.

Up till now our computation of $W_{eff}$ and the equations of
motion has been quite general (modulo the caveats described above),
valid not just near the $n=1$/$n=2$ singularity. However, the
equations of motion \eom\ are clearly quite forbidding, and a
general solution is not readily apparent. Thus it comes as a
rather pleasant surprise that they simplify dramatically at the
$n=1$/$n=2$ singularity! The source of the simplification is the
fact that $r_2=r_3$ at an $n=1$/$n=2$ singularity. (Recall that
the cuts of $y_m$ were chosen to be $(r_1,r_2)$ and $(r_3,r_4)$,
with the ansatz that the $r_i$ are real and $r_1<r_2<r_3<r_4$.)
From \thetaRn, we see this implies that $n=R=0$. Then the elliptic
integrals reduce to $F(\theta|0)=\Pi(0;\theta|0)=\theta$, and the
equations of motion and constraints become:
\eqn\eomsimp{\eqalign{
    & \sqrt{{r_4-r_2\over r_4-r_1}}=\sin\left({\pi\over2}{N_2\over N}\right)\cr
    & r_4-r_1=4\eta\Lambda \cr
    & r_1+2r_2+r_4=0 \cr
    &(r_1-r_4)^2-4r_2(r_1+r_4)=8m^2.\cr
}}
Notice that since $r_2=r_3$, there are four equations for three
unknowns, and thus for generic $m$, $N_1$ and $N_2$ there is no
solution to the equations of motion. In order for there to be a
solution, we must tune the parameter $m$ to special values
depending on $N_1$ and $N_2$. Here $\eta$ is an $2N$th root of
unity, which we must include in order to obtain all of the
$n=1$/$n=2$ singularities. It arises through the presence of the
logarithm in \weffdereval\ and \eom, which should be thought of as
a Veneziano-Yankielowicz-type term \VenezianoAH, in which case it
is more properly written as
$\log\left[\left({r_4+r_3-r_2-r_1\over4\Lambda}\right)^{2N}\right]$.
The equations of motion for $r_2=r_3$ require this logarithm to be
zero, resulting in the $2N$ branches shown in \eomsimp\ labelled
by $\eta$.

Solving \eomsimp\ for $r_i$ and $m$, we obtain
\eqn\singsol{\eqalign{
    & r_1=-\eta\Lambda\left(2+\cos {\pi N_2\over N}\right)\cr
    & r_2=r_3=\eta\Lambda \cos {\pi N_2\over N} \cr
    & r_4=\eta\Lambda\left(2-\cos {\pi N_2\over N}\right)\cr
    & m^2=\eta^2\Lambda^2\left(2+\cos^2 {\pi N_2\over N}\right).\cr
}}
Matching the two parameterizations of the matrix model curve \ymarb, we obtain
the expectation values of the fields $f_0$ and $f_1$ at the double-root singularity:
\eqn\mfexpval{\eqalign{
    &\langle f_0 \rangle = -4\eta^4\Lambda^4\left(1+2\cos^2 {\pi N_2\over N}\right)\cr
    &\langle f_1 \rangle = 8\eta^3\Lambda^3\cos{\pi N_2\over N}.\cr
}}
Using the fact that the total glueball field is related to $f_1$
via $S=-{1\over4}f_1$, we also obtain the expectation value of the
glueball field at the $n=1$/$n=2$ singularity:
\eqn\sexpval{
    \langle S \rangle = -2\eta^3\Lambda^3\cos{\pi N_2\over N}.
}
The general formulas \singsol, \mfexpval\ and \sexpval\ for the
form of the matrix model curve, the parameters of the tree-level
superpotential and the expectation value of the glueball field at
the $n=1$/$n=2$ singularity were not known before. Previously, the
locations of the $n=1$/$n=2$ singularities and the expectation
value of $S$ at the singularities were obtained only for
sufficiently small values of $N$ where the $\CN=2$ factorization
problem could be explicitly solved \refs{\CachazoZK,\FerrariKQ}.
The factorization problem grows in complexity with $N$, but by
studying the glueball equations of motion at the $n=1$/$n=2$
singularity, we have side-stepped the difficulty of the solving
the general factorization problem and obtained the quantum vacua
at the singularity for general $N_1$, $N_2$.

To illustrate the power of our general results, it will be useful
to compare them with an explicit example from \CachazoZK. For the
case of $U(3)$, the only possible choices of breaking pattern with
$n=2$ are $(N_1,N_2)=(1,2)$ and $(2,1)$. \singsol\ and \mfexpval\
then predict that the matrix model curve takes the form
\eqn\ymuthr{
    y_m^2=\left(x^2-m^2\right)^2-4\eta^3\Lambda^3(x+m)
}
with $m=3\eta\Lambda/2$ and $\eta^6=1$. In \CachazoZK, solving the
factorization problem resulted in the matrix model curve
$y_m^2=(x^2-a^2)^2-4\epsilon\Lambda^3(x+a)$, and intersections
with the $n=1$ branch occurred at $8a^3=27\epsilon\Lambda^3$. This
is in precise agreement with the glueball approach.

One can similarly verify that the cases of $U(4)$, $U(5)$ and
$U(6)$ studied in \CachazoZK\ agree exactly with the predictions
of the general formulas \singsol--\sexpval\ derived here. When
performing these checks, and in general when using the formulas of
this section, it is important to remember that $N_1$ and $N_2$ are
defined to be the on-shell periods of the one-form $T(x)$ {\it at
the $n=1$/$n=2$ singularity} (we refer the reader to \CachazoZK\
for the definition of $T(x)$ and the details). This must be
distinguished from the various values of $(N_1,N_2)$ that can be
realized via semiclassical $\Lambda\rightarrow0$ limits on a given
branch of vacua. For instance, in the case of $U(4)$, the
confining branches have unbroken gauge group $U(2)\times U(2)$
semiclassically, while the Coulomb branches have two different
semiclassical limits corresponding to unbroken gauge group
$U(3)\times U(1)$ and $U(2)\times U(2)$. In the general formulas
of this section one must use $(N_1,N_2)=(2,2)$ for the confining
branches and $(N_1,N_2)=(3,1)$ for the Coulomb branches in order
to compare with \CachazoZK, since these are the values of
$(N_1,N_2)$ at the $n=1$/$n=2$ singularities of these
branches.\foot{We thank C.~Ahn for bringing this point to our
attention.}

%

Having solved the equations of motion and obtained the vacua at
the $n=1$/$n=2$ singularity, it is not difficult to take the
calculation off-shell, i.e. expand around these vacua and obtain
the effective glueball superpotential near the singularity. This
can be done, for instance, by expanding the first derivatives
\weffdereval\ of $W_{eff}$ in powers of $f_0-\langle f_0\rangle$
and $f_1-\langle f_1\rangle$ and then integrating. After a
somewhat lengthy calculation, one finds that the effective
glueball superpotential is regular at the singularity when written
in terms of $f_0$ and $f_1$, and all fluctuations are massive.
This agrees with the expectation from the strong-coupling analysis
that the $n=1$/$n=2$ singularity arises solely from a charged
monopole becoming massless, with all other fields massive.

\subsec{The matrix of coupling constants near the singularity}

In order to see from the glueball analysis the effect of the
monopole becoming massless at the singularity, we need to compute
the matrix of gauge couplings $\tau_{ij}$ near the singularity. We
will see that $\tau_{ij}$ tends to zero as an inverse logarithm as
we approach the singularity. This will indicate that an additional
charged field is becoming massless. Notice, however, that the
glueball description fails to provide an explanation for this
massless, charged field. For that, we need the strong-coupling
description, in which this new massless field is a monopole.
Moving along the $n=2$ branch, the monopole acquires a bare mass,
while moving along the $n=1$ branch condenses the monopole,
spontaneously breaking $U(1)^2$ down to $U(1)$ via the Higgs
mechanism.

The matrix of gauge couplings is given by the formula \CachazoRY:
\eqn\tauij{
    {1\over2\pi i}\tau_{ij}=
    {\partial \Pi_i \over \partial S_j}-\delta_{ij}{1\over N_i}\sum_{l=1}^{n} N_l{\partial \Pi_i \over \partial S_l}.
}
For $n=2$, this simplifies to
\eqn\tauijmat{
    \tau_{ij}=\left(\matrix{ \tau_{11} & \tau_{12} \cr
                             \tau_{21} & \tau_{22} }\right)
    = 2\pi i {\partial \Pi_2 \over \partial S_1} \left(\matrix{ -{N_2\over N_1} & 1 \cr
                             1 & -{N_1\over N_2}}\right),
}
where we have used the fact that ${\partial \Pi_1 \over \partial
S_2}={\partial\Pi_2\over\partial S_1}$ since $\Pi_i$ is itself a
derivative ${\partial \CF \over \partial S_i}$ of the
prepotential. Thus we see that the matrix of coupling constants
depends only on the single derivative ${\partial
\Pi_2\over\partial S_1}$, evaluated the extremum of the effective
superpotential. In appendix A we calculate this partial
derivative; it can be related to the derivatives with respect to
$f_0$ and $f_1$ via the chain rule. The answer, to leading order
in $R$, is simply
\eqn\partialfin{
    {\partial \Pi_2 \over \partial S_1}=
    {\pi i N_1 N_2\over N^2}{1\over \log\left({16\over R}\right)}+\CO(R).
}
As we approach the singularity, we see that $\tau_{ij}$ has the
limiting behavior:
\eqn\tauijrew{\eqalign{
    {1\over 2\pi i}\tau_{ij}&= {\pi i N_1 N_2\over N^2}{1\over \log\left({16\over R}\right)}
    \left(\matrix{ -{N_2\over N_1} & 1 \cr 1 & -{N_1\over N_2}}\right)+\CO(R).
}}
Therefore $\tau_{ij}\rightarrow 0$ as an inverse logarithm as we
approach the $n=1$/$n=2$ singularity. This means that $\tau_{ij}$
is continuous as we pass from the $n=2$ branch to the $n=1$
branch, since $\tau_{ij}=0$ trivially on the $n=1$ branch. More
importantly, however, the logarithm indicates that the gauge
coupling constant of the non-trivial $U(1)$, given roughly by $1/\tau$, diverges as we
approach the singularity. The divergence is due to an additional
monopole becoming massless at the singularity, which causes an IR
divergence in the renormalized coupling of the $U(1)$ under which it is charged.

\newsec{$\CN=1$ Argyres-Douglas points}

The third and final set of singularities we will consider are the
$\CN=1$ AD points
\refs{\ArgyresJJ\ArgyresXN\EguchiVU\TerashimaPN-\TerashimaZV}.
Such points are believed to represent a novel class of interacting
$\CN=1$ SCFTs at low-energies. They arise when a tree-level
superpotential for $\Phi$ lifts the flat directions of $\CN=2$
moduli space, leaving vacua that include mutually non-local
monopoles as massless degrees of freedom. The conditions for a
tree-level superpotential to lead to an $\CN=1$ AD point include
an $\CN=2$ curve with triple or higher-order roots and a matrix
model curve that vanishes at these roots.\foot{As noted in
\EguchiWV, this second condition is necessary in order for the
non-local monopoles to remain uncondensed. A non-zero monopole
condensate will produce a mass gap in the $U(1)$ under which it is
charged.} Recall that the tree-level superpotential determines the
$\CN=2$ and matrix model curves via the solution of the
factorization problem \refs{\CachazoJY, \CachazoZK}:
\eqn\factorize{\eqalign{
    y^2&=P_N(x)^2-4\Lambda^{2N}=F_{2n}(x)H_{N-n}(x)^2\cr
    y_m^2&=W'_k(x)^2+f_{k-1}(x)=F_{2n}(x)Q_{k-n}(x)^2.\cr
}}
For simplicity, we shall assume that the degree $k+1$ of the
tree-level superpotential satisfies $k<N$, so that the polynomials
$H_{N-n}$ and $Q_{k-n}$ are independent.

Notice that the $\CN=2$ curve is comprised of two $N$th-degree
polynomial factors $P_N(x)\pm 2\Lambda^N$ that cannot share any
common roots. Thus the $\CN=2$ curve can have at most an $N$th
order root, at which $P_N(x)=x^N\pm 2\Lambda^N$. We will restrict
our attention to the $\CN=1$ singularities that can be obtained
from these maximally singular points, as this will simplify our
calculations considerably. At the same time, we do not expect to
lose too much in the way of physics with such a restriction, since
for a given $N$ the maximal AD points are the locations in moduli
space where the largest number ($N-1$) of mutually non-local,
linearly independent monopoles become simultaneously massless.
These $N-1$ monopoles are charged under $\left[{N\over2}\right]$
of the $U(1)$ factors \ArgyresJJ.

To understand the relationship between the $\CN=1$ and the $\CN=2$
AD points, it helps to study in greater detail the factorization
problem \factorize. At the maximal AD point,
$P_N(x)=x^N-2\Lambda^N$ (we choose the minus sign without loss of
generality) and the $\CN=2$ curve takes the form
\eqn\swcurvemax{
    y^2=x^N(x^N-4\Lambda^N).
}
The $\CN=2$ curve must factorize into the
polynomials $F_{2n}(x)$ and $H_{N-n}(x)^2$. Thus, for a given $N$
and $n$, we must have
\eqn\swmaxfac{\eqalign{
    &H_{N-n}(x)=x^{N-n} \cr
    &F_{2n}(x)=x^{2n-N}(x^N-4\Lambda^N),\cr
}}
implying the following form for the matrix model curve:
\eqn\mmmax{
    y_m^2=F_{2n}(x)Q_{k-n}(x)^2=\left(x^n Q_{k-n}(x)\right)^2-4\Lambda^N
                          x^{2n-N}Q_{k-n}(x)^2.
}
Recall that we are assuming $k<N$; thus the second term is a
polynomial of degree $< k$. Therefore we have
\eqn\mmfac{\eqalign{
    &W'_k(x)=x^n Q_{k-n}(x)\cr
    &f_{k-1}(x)=-4\Lambda^N x^{2n-N}Q_{k-n}(x)^2.\cr
}}
Together, \swmaxfac\ and \mmfac\ represent the complete solution
to the factorization problem at the maximal AD point. We see that
the maximal AD point actually leads to a number of continuous
families of $\CN=1$ vacua, obtained by varying the parameters of
$Q_{k-n}(x)$. These continuous families are indexed by the
integers $(k,n)$, which must satisfy the inequalities
\eqn\knineq{
    \left[{N+1\over2}\right] \le n \le k < N.
}
To avoid overcounting vacua, we will assume that $Q_{k-n}(x)$ is
non-zero at $x=0$.

Not all of these $\CN=1$ vacua will to lead to different
interacting CFTs in the IR, however. Rather, we expect that many
of the details of these vacua will become irrelevant in the IR,
leading to a much smaller, discrete set of universality classes of
$\CN=1$ SCFTs. The simplest possibility is that the features of
the IR theory are determined by the most relevant term of the
tree-level superpotential. Let us write
\eqn\Qpoly{
    Q_{k-n}(x)=\sum_{i=0}^{k-n} q_i x^i
}
with $q_{k-n}=1$ and $q_0\ne 0$. Then using \mmfac\ to relate
$W(x)$ to $Q_{k-n}(x)$, we see that the most relevant (i.e.
lowest-order) term in the tree-level superpotential is
\eqn\wtreerel{
    W(\Phi)={q_0\over n+1}\, \Tr\,\Phi^{n+1}+\CO(\Tr\,\Phi^{n+2}).
}
As we flow down to the IR, we expect the operator $\Tr\,
\Phi^{n+1}$ to become marginal or irrelevant, and the higher order
terms in $W(\Phi)$ to become irrelevant. This suggests that the IR
physics depends only on the degree of singularity $n+1$ at $x=0$
of the tree-level superpotential, and not on its overall degree
$k+1$. We are thus led to propose that the universality classes
are labelled by $n$ alone. Let us call these universality classes
$AD_n$ for want of a better name.

We can adduce some evidence for our proposal from the recent work
of \EguchiWV. There, an attempt was made to define scaling
operators and calculate their dimensions for the $\CN=1$ $AD$
points obtained with monomial superpotential $W'(x)=x^n$. The
analysis was restricted to the range $n> n_{min}$, and it was
found that the scaling dimensions of appropriately defined chiral
operators depended only on the behavior of the matrix model curve
very near $x=0$. From \mmmax, we see that near $x=0$ the matrix
model curve takes the form
\eqn\mmxzero{
y_m^2\approx -4\Lambda^Nq_0 x^{2n-N}
}
and in particular is independent of the overall degree $k$ of the
superpotential. Thus, our proposal that the $\CN=1$ AD points
depend only on $n$, the degree of the most relevant term in the
superpotential, and not on the overall degree $k$, is consistent
with the approach of \EguchiWV.

\topinsert
    \centerline{\relabelbox\epsfxsize=0.75\hsize\epsfbox{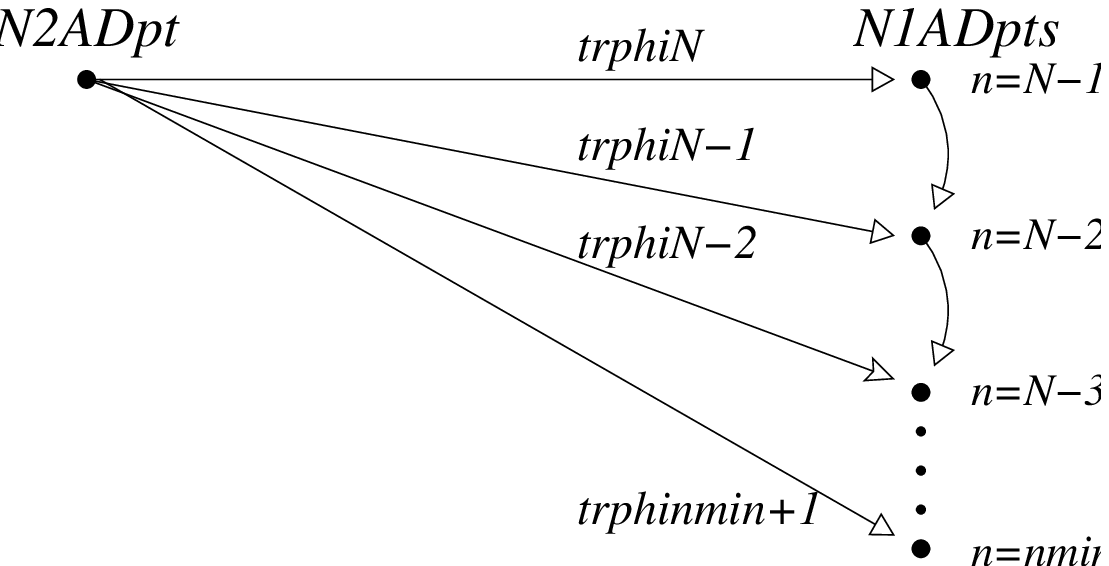}
    \relabel {N2ADpt}{$\CN=2$ AD point}
    \relabel {N1ADpts}{$\CN=1$ AD points}
    \relabel {trphiN}{$\Tr\,\Phi^N$}
    \relabel {trphiN-1}{$\Tr\,\Phi^{N-1}$}
    \relabel {trphiN-2}{$\Tr\,\Phi^{N-2}$}
    \relabel {trphinmin+1}{$\Tr\,\Phi^{n_{min}+1}$}
    \relabel {n=N-1}{$n=N-1$}
    \relabel {n=N-2}{$n=N-2$}
    \relabel {n=N-3}{$n=N-3$}
    \relabel {n=nmin}{$n=n_{min}$}
    \endrelabelbox}
    \noindent{\ninepoint\sl\baselineskip=8pt {\bf Figure 1}:
    {\sl $\;$ A schematic depiction of the proposed renormalization group flow between AD points. As
    usual, the arrows indicate flow towards the IR. Starting from the
    $\CN=2$ AD point (no tree-level superpotential), we can flow to a
    particular $\CN=1$ $AD_n$ point by perturbing with the operator
    $\Tr\,\Phi^{n+1}$. Similarly, starting from an $AD_n$ point, we can
    flow to an $AD_m$ point with $m<n$ by perturbing with
    $\Tr\,\Phi^{m+1}$. The RG cascade must terminate at $n=n_{min}$;
    perturbing further with relevant operators will send the theory to
    a trivial IR fixed point.
    }}
    \bigskip
\endinsert

If our proposal is correct, we are led to a satisfying picture of
the IR renormalization group flow as a cascade-like descent to
CFTs of smaller and smaller $n$, as shown in figure 1. 
Starting from the $\CN=2$ maximal AD point with $W(x)=0$ we can
perturb by a monomial superpotential $W'(x)=x^n$, inducing RG flow
to the $\CN=1$ $AD_n$ point. Perturbing further by a monomial term
of lower degree $W'(x)\rightarrow W'(x)+x^{n-1}$ takes us to the
$AD_{n-1}$ point. In this manner, we can descend from the $\CN=2$
AD point down successively through the $\CN=1$ $AD_n$ points via
superpotential perturbations. The descent must end at the minimal
value of $n_{min}=\left[{N+1\over 2}\right]$; perturbing by a
monomial of lower degree will send us to a trivial free theory in
the IR.

There has been some debate as to whether the minimal value
$n_{min}$ corresponds to an interacting or a free IR theory.
Although the authors of \EguchiWV\ restricted their analysis to
$n>n_{min}$, they suggested that the IR theory was trivial and
free for the borderline case $n=n_{min}$. For $N$ even, this
suggestion is almost certainly correct. As pointed out in
\EguchiWV, the monopole condensate corresponding to a given double
root $p$ of the $\CN=2$ curve is proportional to the matrix model
curve at $p$ \deBoerAP:
\eqn\mplcond{
    \langle q \tilde{q}\rangle \propto y_m(x=p).
}
When $N$ is even, the matrix model curve is non-zero at $x=0$, and
therefore the monopoles remain condensed even at the AD point.
Thus the theory is indeed massive and trivial.

On the other hand, $y_m=0$ at $x=0$ for $N$ odd, and therefore the
monopole condensates and the mass gap vanish at the AD point. Note
that $N=3$, $n_{min}=2$ is the original $\CN=1$ AD point discussed
in the work of Argyres and Douglas. The vanishing of the monopole
condensates for the original $\CN=1$ $AD$ point was also
demonstrated explicitly in \GorskyEJ. But the authors of
\EguchiWV\ argued that the IR theory for $n=n_{min}$ and $N$ odd
is nevertheless trivial, because the matrix model curve is
non-singular at $x=0$. While this fact does not necessarily
signify anything {\it per se}, it does imply that the chiral
operators defined in \EguchiWV\ do not exhibit scaling behavior.
Thus there is no evidence for an interacting IR theory in that
approach. However we wish to argue in the rest of this section
that there does in fact exist evidence for an interacting theory
when $N$ is odd. By studying both the effective superpotential in
both the glueball and the strong-coupling descriptions, we will
show that the effective descriptions look dramatically different
for $N$ even and $N$ odd. We will confirm that IR theory is indeed
trivial for $N$ even, but seems to be non-trivial for $N$ odd.

\subsec{The effective glueball superpotential}

We start with the case of $N$ odd. The matrix model curve for $N$
odd and $n=n_{min}$ is
\eqn\mmnodd{
y_m^2=x(x^N-4\Lambda^N)+g_{n-1}(x),
}
where we are using $g_{n-1}(x)$ to denote the off-shell
fluctuations of the matrix model curve. The $n$ coefficients of
$g_{n-1}(x)$ are related to the glueball fields through the
periods of $y_m$. Notice that we have assumed $k=n$, i.e. a
monomial superpotential. We will continue to assume this for the
rest of the section. If the IR physics is determined by $n$ alone,
as we have proposed, then we lose no generality in making this
simplifying assumption.

We wish to compute the effective glueball superpotential $W_{eff}$
derived from this matrix model curve. While we will not be able to
obtain an explicit formula for $W_{eff}$ for general $N$, we will
be able to constrain the form of $W_{eff}$ using a discrete ${\bf
Z}_N$ symmetry. This alone will allow some surprising conclusions
to be drawn about the low-energy theory. The ${\bf Z}_N$ symmetry
is closely related to the $U(1)_R$ symmetry of the microscopic
theory. Let us now take a moment to describe in detail how the
${\bf Z}_N$ symmetry comes about. Here we will give only heuristic
arguments for the existence of such a symmetry; we refer the
reader to appendix B 
for a more detailed derivation using matrix model techniques.

The microscopic theory has a $U(1)_R$ symmetry which is unbroken
by the monomial tree-level superpotential
$W(\Phi)=\Tr\,\Phi^{n+1}$, provided we give $\Phi$ an $R$-charge
${2\over n+1}$. Quantum mechanically, $P_N(x)=x^N-2\Lambda^N$ at
the AD point, and thus $\langle\Tr\,\Phi^j\rangle=0$ for
$j=1,\dots,N-1$ and $\langle\Tr\,\Phi^N\rangle=2\Lambda^N$. As a
result, the $U(1)_R$ symmetry is spontaneously broken down to
${\bf Z}_N$ by the expectation value of $\Tr\,\Phi^N$. Under the
residual ${\bf Z}_N$ symmetry, $\Phi$ and the superpotential can
be chosen to transform as
\eqn\phiwzn{
    \Phi\rightarrow e^{{4\pi i \over N}}\Phi,
    \qquad W(\Phi)\rightarrow e^{6\pi i \over N}W(\Phi).
}
Now consider integrating out $\Phi$ and passing to the low-energy
effective description in terms of the glueball fields. Since the
residual ${\bf Z}_N$ symmetry is unbroken in the exact quantum
theory, the effective superpotential should also transform as
\eqn\weffznNodd{
    W_{eff}\rightarrow e^{6\pi i\over N}W_{eff}
}
under the action of ${\bf Z}_N$. (This can be shown more
rigorously through explicit matrix model computations; see
appendix B.)

It remains to determine how the glueball fields of the low-energy
theory, or equivalently the coefficients of $g_{n-1}(x)$,
transform under ${\bf Z}_N$. But this is not hard to see. The
on-shell matrix model curve is $y_m^2=x(x^N-4\Lambda^N)$; if we
make the natural identification of the $R$-charge of $x$ with the
$R$-charge of $\Phi$, it follows that $y_m^2$ has charge
${4\pi\over N}$ under ${\bf Z}_N$. This will also be true
off-shell if the coefficients of $g_{n-1}(x)=\sum_{j=0}^{n-1} g_j
x^j$ transform as
\eqn\fjznNodd{
    g_j \rightarrow e^{4\pi i\,(1-j)/N}g_j.
}

Together, \weffznNodd\ and \fjznNodd\ describe the action of the
${\bf Z}_N$ symmetry on the low-energy theory. We are now ready to
understand the implications of this discrete symmetry on the
effective superpotential. Recalling that $W_{eff}$ must be
holomorphic in the fields $g_j$, we find that the ${\bf Z}_N$
symmetry severely restricts the form of the effective
superpotential:
\eqn\weffNodd{
W_{eff}=g_{n-1}
G\left(g_0^{N},g_1,g_2g_0,\dots,g_{n-1}g_0^{n-2}\right)
}
for some holomorphic function $G$. While this does not determine
the full form of $W_{eff}$, it does allow us to write down the
linear and quadratic terms up to undetermined constants:
\eqn\weffquadNodd{
W_{eff}\approx c_0 g_{n-1}+\sum_{i=1}^{\left[{n\over2}\right]} c_i
g_i g_{n-i}+\dots
}
The linear term can actually be eliminated, because if $c_0\ne 0$,
this merely tells us that $\langle g_1 \rangle \approx -c_0/c_1
\ne 0$. But from the form of the matrix model curve \mmnodd, we
deduce that giving a vev to $g_1$ is indistinguishable from
redefining $\Lambda$ by a constant. Therefore we might as well
redefine $\Lambda$ so that $c_1=0$.

No other terms are possible at quadratic order in the fields. In
particular, a mass term for $g_0$ is forbidden by the ${\bf Z}_N$
symmetry! Mass terms are clearly allowed for the other $n-1$
fields. Therefore, in the absence of additional, accidental
symmetries (and assuming regularity of the K\"ahler potential) we
except there to be exactly one massless glueball in the low-energy
spectrum for $N$ odd and $n=n_{min}$.

Now let us consider the case of $N$ even. The matrix model curve
for $N$ even and $n=n_{min}$ is
\eqn\mmneven{
y_m^2=(x^N-4\Lambda^N)+g_{n-1}(x).
}
The arguments for the ${\bf Z}_N$ symmetry are nearly identical to
those given above for $N$ odd. Therefore we save the details for
the appendix and merely summarize the results here. We once again
have $x\rightarrow e^{4\pi i\over N} x$, but now
\eqn\weffznNeven{
    W_{eff}\rightarrow e^{4\pi i \over N} W_{eff}
}
and
\eqn\fjznNeven{
    g_j \rightarrow e^{-{4\pi i\, j\over N}}g_j.
}
This restricts the form of the effective superpotential to be
\eqn\weffformNeven{
W_{eff}=g_{n-1}
F(g_0,g_1g_{n-1},\dots,g_{n-2}g_{n-1}^{n-2},g_{n-1}^{n})
}
for some holomorphic function $F$. To quadratic order, the
possible terms are therefore
\eqn\weffquadNeven{
    W_{eff}\approx c_0g_{n-1}+\sum_{i=0}^{\left[{n-1\over2}\right]}c_i g_i g_{n-1-i}.
}
Again the linear term can be eliminated by a shift in $\Lambda$.
However, in contrast to the odd $N$ case, no mass terms are
forbidden for the glueball fields for even $N$. In the absence of
additional symmetries, we expect all $c_i\ne 0$. This is evidence
that for $N$ even and $n=n_{min}$ all the fluctuations are massive
around the AD point. This is quite different from the case of $N$
odd!

\subsec{The strong coupling superpotential}

In this subsection, we will analyze the effective superpotential
in the strong-coupling description, where we treat the tree-level
superpotential as a small perturbation of the underlying $\CN=2$
theory. In the strong coupling analysis, the low-energy degrees of
freedom are the fields $u_r={1\over r}\Tr\, \Phi^r$ and some
number of light monopole fields $q_l$, $\tilde{q}_l$,
$l=1,\dots,N-n$. Their interactions are described by the effective
superpotential
\eqn\weffntwo{
    W_{eff}=W(u)+\sum_{l=1}^{N-n} M_{l}(u) q_l \tilde{q}_l,
}
where $W(u)$ is the tree-level superpotential and $M_l(u)$ are the
monopole masses \SeibergRS. Integrating out the monopole fields
implies $M_l(u)=0$ and leads to an effective superpotential for
the fields $u_r$, $r=1,\dots,n$ alone. For the monomial
superpotential under consideration, the effective superpotential
after integrating out the monopole fields is simply
\eqn\weffmon{
    W_{eff}=u_{n+1}(u_1,\dots,u_n),
}
where the $N-n$ equations $M_l(u)=0$ are used to re-express
$u_{n+j}$ in terms of the fields $u_1,\dots,u_n$. Determining the
function $u_{n+1}(u_1,\dots,u_n)$ will be the focus of the
remainder of our strong-coupling analysis.

In practice, it is very difficult to determine the function
$u_{n+1}(u_1,\dots,u_n)$ directly from the massless monopole
equations. Fortunately, there exists an easier way, namely the
factorization of the $\CN=2$ curve. Requiring there to be $N-n$
massless monopoles is equivalent to requiring the $\CN=2$ curve to
factorize into $N-n$ double roots \HananyNA:
\eqn\swfactor{
    y^2=P_N(x)^2-4\Lambda^{2N}=F_{2n}(x)H_{N-n}(x)^2.
}
If we write
\eqn\pnsum{
    P_N(x)=\sum_{k=0}^{N} s_k x^{N-k}
}
with $s_0=1$, then the coefficients of $P_N(x)$ are simply related
to the fields $u_r$ through the equations
\eqn\pncoeff{
k s_k+\sum_{r=1}^{k}r u_r s_{k-r}.
}
The factorization \swfactor\ amounts to a system of $2N$ equations
for $2N+n$ unknowns (the coefficients of $P_N$, $F_{2n}$ and
$H_{N-n}$), and thus only $n$ are independent. Using \pncoeff, we
can express all of the coefficients in terms of $u_1,\dots,u_n$,
and this will produce the desired dependence of $u_{n+j}$ on
$u_1,\dots,u_n$.

We can simplify the problem even further in the case of the
maximal AD point. At the maximal AD point,
$P_N(x)\big|_{AD}=x^N-2\Lambda^N$ and
\eqn\swadpt{
    y^2\big|_{AD}=(P_N(x)+2\Lambda^N)(P_N(x)-2\Lambda^N)\big|_{AD}=x^N(x^N-4\Lambda^N).
}
The roots in the second factor of \swadpt\ are all separated by
$\CO(\Lambda)$, so perturbing $\langle u_i\rangle_{AD}\rightarrow
\langle u_i\rangle_{AD}+\delta u_i$ slightly away from the AD
point on a branch with $N-n$ double roots must leave all the
double roots in the first factor. Therefore we have reduced the
factorization problem from a system of $2N$ equations to a system
of $N$ equations:
\eqn\swfacsimp{
    P_N(x)+2\Lambda^N=\cases{H_{N-n}(x)^2 & $N$ even\cr
                            (x-a)H_{N-n}(x)^2 & $N$ odd\cr}
}

For $N$ even, the factorization problem can be neatly solved as a
power series in $u_1,\dots,u_n$. The resulting effective
superpotential is given to quadratic order by
\eqn\weffNevenur{
W_{eff}={1\over4}\sum_{r=1}^{n}u_r u_{n+1-r} +\CO(u^3).
}
Thus the strong-coupling effective superpotential \weffNevenur\
for $N$ even is regular at the AD point, and all the fluctuations
about the vacuum are massive.

We have skipped the full proof of \weffNevenur, because while
straightforward, it is messy and not very illuminating. But
perhaps it will help to illustrate the general idea by working out
the simplest example of $N=4$. In this case, $n=2$, so we are
interested in finding $u_3(u_1,u_2)$ by solving \swfacsimp:
\eqn\swfacNfour{
    P_4(x)+2\Lambda^4=H_2(x)^2=(x^2+c_1 x+c_2)^2.
}
Substituting \pnsum\ for $P_4(x)$ and equating the coefficients of
the two sides of \swfacNfour\ immediately leads to
\eqn\csol{
c_1={1\over2}s_1,\quad
c_2={1\over2}\left(s_2-{1\over4}s_1^2\right),
}
and so
\eqn\sthr{
    s_3=2c_1c_2={1\over2}s_1\left(s_2-{1\over4}s_1^2\right).
}
It is simple enough now to use \pncoeff\ to translate \sthr\ into
a formula for $u_3$. The result is
\eqn\uthr{
    u_3={1\over2}u_1u_2-{1\over24}u_1^3,
}
which verifies the general result \weffNevenur\ for $N=4$.

Unfortunately the story is not so simple for $N$ odd. In general,
$u_{n+1}$ will be determined from an $n(n+1)$ degree
quasi-homogeneous polynomial equation
\eqn\unNodd{
    R_{n(n+1)}(u_1,\dots,u_{n+1})=0.
}
Thus the complexity of the solution grows rapidly with increasing
$N$. One can simplify things somewhat by restricting the analysis
to $SU(N)$; this amounts to setting $u_1=0$. The trace can always
be restored at the end of the calculation by translating the
$SU(N)$ variables back into the $U(N)$ variables. 
Even with this simplification, however, we are not able to obtain
a general formula for the polynomial $R_{n(n+1)}$. To illustrate
the rapidly growing complexity of these polynomials, we display
here the results for $SU(3)$ and $SU(5)$:
\eqn\Rex{\eqalign{
    & R_6(u_2,u_3)=27 u_3^2-4u_2^3\cr
    & R_{12}(u_2,u_3,u_4)=1600 u_4^3-1360
    u_2^2u_4^2+(384u_2^4-360u_2u_3^2)u_4-(36u_2^6-92u_2^3u_3^2+135u_3^4).\cr
}}

Although for $N$ odd we lack an explicit formula for $W_{eff}$, we
can make a few general statements. Because of the presence of
higher powers of $u_{n+1}$, the solution $u_{n+1}(u_1,\dots,u_n)$
of $R_{n(n+1)}=0$ will in general be non-analytic at
$u_1=\dots=u_n=0$. In particular, the second derivatives of
$W_{eff}$ will generally be divergent at the maximal AD point. For
instance, the effective superpotential for $SU(3)$ is
\eqn\QNthree{
    W_{eff}=u_3(u_2)=\sqrt{{4\over27}}\,u_2^{3/2},
}
and the field $u_2$ is seen to have a divergent mass term at
$u_2=0$. (The formula for $U(3)$ will be somewhat more
complicated, although the conclusions are unchanged.) Thus for $N$
odd, the strong-coupling analysis produces rather pathological
singularities in the effective superpotential. Such pathologies in
the effective superpotential generally indicate a breakdown of the
effective description and signal the presence of extra massless
fields. Contrast this with the case of $N$ even, where both the
glueball analysis and the strong-coupling analysis resulted in
perfectly regular superpotentials with all fluctuations massive.
This suggests that the additional massless fields are precisely
the mutually non-local monopole fields that remain uncondensed at
the AD point for $N$ odd. We take this to be
evidence that the $\CN=1$ AD point with $n=n_{min}$ and $N$ odd is indeed
a non-trivial, interacting CFT at low-energies.

\vskip 1cm
\centerline{\bf Acknowledgements}

First and foremost I would like to thank my advisor N. Seiberg for
suggesting this problem and for his constant guidance and support
throughout every stage of this work. It is also a pleasure to
thank F.~Cachazo, S.~Murthy, M.~Douglas, K.~Intriligator and
M.~Strassler for useful discussions. This work was supported
by an NSF Graduate Research Fellowship. Any opinions,
findings, and conclusions or recommendations expressed in this
material are those of the author and do not necessarily reflect
the views of the National Science Foundation.

\appendix{A}{Calculating derivatives of periods on the $n=2$ branch}

In this appendix, we evaluate the derivatives of the periods $S_i$
and $\Pi_i$ of the matrix model curve on the $n=2$ branch, with
cubic tree-level superpotential. The periods of interest are the
glueball fields $S_i$:
\eqn\siper{\eqalign{
    & 2\pi i S_1=\int_{r_1}^{r_2}\sqrt{(x-r_1)(x-r_2)(x-r_3)(x-r_4)}\,dx\cr
    & 2\pi i S_2=\int_{r_3}^{r_4}\sqrt{(x-r_1)(x-r_2)(x-r_3)(x-r_4)}\,dx\cr
}}
and their conjugate periods:
\eqn\piper{\eqalign{
    & 2\pi i \Pi_1=\int_{-\Lambda_0}^{r_1}\sqrt{(x-r_1)(x-r_2)(x-r_3)(x-r_4)}\,dx\cr
    & 2\pi i \Pi_2=\int_{\Lambda_0}^{r_4}\sqrt{(x-r_1)(x-r_2)(x-r_3)(x-r_4)}\,dx.\cr
}}
These periods are combined to give the effective glueball
superpotential:
\eqn\weffdef{
    W_{eff}=2\pi i\,(N_1\Pi_1+N_2\Pi_2+b_2 S_2)+2 N S\log\left({\Lambda\over\Lambda_0}\right).
}
Let us start with the derivatives of $\Pi_i$:
\eqn\pider{\eqalign{
    4\pi i{\partial \Pi_1\over \partial f_j} &=
    \int_{-\Lambda_0}^{r_1}{x^j\over\sqrt{(x-r_1)(x-r_2)(x-r_3)(x-r_4)}}\,dx \cr
    4\pi i{\partial \Pi_2\over \partial f_j} &=
    -\int_{r_4}^{\Lambda_0}{x^j\over\sqrt{(x-r_1)(x-r_2)(x-r_3)(x-r_4)}}\,dx. \cr
}}
With some effort, these integrals can be evaluated and expressed
in terms of elliptic integrals of the first and third kinds:
\eqn\pidereval{\eqalign{
4\pi i{\partial \Pi_1\over\partial
f_0}&={2\over\sqrt{(r_4-r_2)(r_3-r_1)}}F(\theta | R)\cr 4\pi
i{\partial \Pi_1\over\partial
f_1}&=\log\left({r_4+r_3-r_2-r_1\over 4\Lambda_0}\right)
+{2\left(r_1\,F(\theta | R)+(r_2-r_1)\,\Pi(n;\theta |
R)\right)\over\sqrt{(r_4-r_2)(r_3-r_1)}}\cr 4\pi i{\partial
\Pi_2\over\partial f_0}&={2\over\sqrt{(r_4-r_2)(r_3-r_1)}}\Delta
F_2(r_i)\cr 4\pi i{\partial \Pi_2\over\partial
f_1}&=\log\left({r_4+r_3-r_2-r_1\over 4\Lambda_0}\right)
+{2\left(r_1\,\Delta
F_2(r_i)+(r_2-r_1)\,\Delta\Pi_2(r_i)\right)\over\sqrt{(r_4-r_2)(r_3-r_1)}}\cr
}}
where
\eqn\deltafpitwo{\eqalign{
    &\Delta F_2(r_i)=F(\theta | R)-F({\pi\over2} | R) \cr
    &\Delta\Pi_2(r_i)=\Pi(n; \theta | R)-\Pi(n; {\pi\over2}|R)\cr
    &\sin^2\theta = {r_4-r_2\over r_4-r_1},\quad R={(r_3-r_2)(r_4-r_1)\over
        (r_3-r_1)(r_4-r_2)},\quad n={r_3-r_2\over r_3-r_1}.
}}
The formulas in \pidereval\ are exact, up to terms of
$\CO(1/\Lambda_0)$ which are irrelevant in the
$\Lambda_0\rightarrow\infty$ limit. Here $F$ and $\Pi$ are the
elliptic integrals of the first and third kinds, respectively. We
recall their definitions:
\eqn\elliptic{\eqalign{
    & F(\theta | R)=\int_{0}^{\theta} {d\theta'\over \sqrt{1-R \sin^2\theta'}} \cr
    & \Pi(n; \theta | R) = \int_{0}^{\theta} {d\theta'\over (1-n \sin^2\theta')\sqrt{1-R \sin^2\theta'}}.\cr
}}
Using \pidereval\ in the formula \weffdef\ for the effective
glueball superpotential immediately results in the glueball
equations of motion \weffdereval\ (for $b_2=0$).

Now let us turn to the derivatives of the $S_i$. It actually
suffices to consider only the derivatives of $S_2$, since
$S_1+S_2=S=-{1\over4}f_1$:
\eqn\sder{\eqalign{
4\pi i{\partial S_2\over \partial f_0} &= -i\int_{r_3}^{r_4}
{1\over\sqrt{(x-r_1)(x-r_2)(x-r_3)(r_4-x)}}\,dx \cr 4\pi
i{\partial S_2\over \partial f_1} &= -i\int_{r_3}^{r_4}
{x\over\sqrt{(x-r_1)(x-r_2)(x-r_3)(r_4-x)}}\,dx. \cr
}}
Note that we have implicitly chosen the branch of the square root
in the denominator by pulling out a factor of $-i$ in front of the
integral. One can verify that this is the correct branch of the
square root to choose by computing $S_1$ and making sure that they
indeed combine to give $-{1\over4}f_1$. The integrals \sder\
evaluate to:
\eqn\stwodereval{\eqalign{
4\pi i{\partial S_2\over\partial
f_0}&=-{2i\over\sqrt{(r_4-r_2)(r_3-r_1)}}F({\pi\over2}| R_2)\cr
4\pi i{\partial S_2\over\partial f_1}&=
-{2i\over\sqrt{(r_4-r_2)(r_3-r_1)}}\left(r_2\,F({\pi\over2}|R_2)
+(r_3-r_2)\,\Pi(n_2;{\pi\over2}|R_2)\right)\cr
}}
where $R_2$ and  $n_2$ are given by
\eqn\nrtwodef{
R_2 = 1-R={(r_2-r_1)(r_4-r_3)\over (r_4-r_2)(r_3-r_1)},\quad
n_2={1-R\over1-n}={r_4-r_3\over r_4-r_2}.
}

Finally, we will compute the partial derivative ${\partial
\Pi_2\over\partial S_1}$ near the $n=1$/$n=2$ singularity. This
calculation is relevant for obtaining the matrix of coupling
constants at the singularity. First, we use the chain rule to
relate the derivative with respect to $S_1$ to derivatives with
respect to $f_0$ and $f_1$:
\eqn\chain{\eqalign{
    {\partial \Pi_2 \over \partial S_1}
    &=
    {\partial \Pi_2 \over \partial f_1}{\partial f_1 \over \partial S_1}
    +{\partial \Pi_2 \over \partial f_0}{\partial f_0 \over \partial
    S_1} \cr
    &=
    4 \left(\kappa\, {\partial \Pi_2 \over \partial f_0}-{\partial \Pi_2 \over \partial f_1}\right).
}}
Here we have defined the ratio $\kappa = {\partial S_2 \over
\partial f_1}/{\partial S_2\over \partial f_0}$, and we have used the fact that
$S_1=S-S_2=-4f_1-S_2$ to eliminate $S_1$. The ratio $\kappa$ is
the source of the logarithmic singularity. Using \stwodereval\
gives:
\eqn\kappadef{
\kappa = r_2+{(r_3-r_2)\Pi(n_2;{\pi\over2}|R_2)\over F({\pi\over
2}| R_2)}.
}
As $r_2-r_3\rightarrow0$, we see from \nrtwodef\ that $n_2$ and
$R_2\rightarrow 1$. The limiting behavior of $F({\pi\over2}| R_2)$
as $R_2\rightarrow 1$ is
\eqn\flimit{
F({\pi\over2}|R_2)={1\over2}\log\left({16\over
1-R_2}\right)+\CO(1-R_2)={1\over2}\log\left({16\over
R}\right)+\CO(R).
}
On the other hand, the limiting behavior of
$(r_3-r_2)\Pi(n_2;{\pi\over2}|R_2)$ is finite, as one can see by
returning to the integral definition \elliptic\ of $\Pi$. Let us
write $1/n_2=1+\epsilon$ and $1/R_2=1+\alpha\,\epsilon$, where
$\epsilon\rightarrow 0$ and $\alpha$ is some constant. Then from
\elliptic, we obtain
\eqn\pilimit{\eqalign{
    \Pi(n_2;{\pi\over2}|R_2)
    &=
    {1\over n_2\sqrt{R_2}} \int_{0}^{1} {dt\over (t^2+\epsilon)\sqrt{(t^2+\alpha\,\epsilon)(1-t^2)}}\cr
    &=
    {1\over \epsilon\,n_2\sqrt{R_2}}\int_{0}^{1\over\sqrt{\epsilon}}
    {dt \over(t^2+1)\sqrt{(t^2+\alpha)(1-\epsilon\,t^2)}}\cr
    &=
    {r_4-r_2\over r_3-r_2}\sqrt{{r_3-r_1\over r_4-r_3}}\,\theta+\CO(\epsilon^0)\cr
}}
where $\theta$ is defined in \thetaRn. Combining \flimit\ and
\pilimit, we see that
\eqn\kapparew{
\kappa = r_2+{2(r_4-r_2)\sqrt{{r_3-r_1\over r_4-r_3}}\,\theta\over
\log\left({16\over R}\right)}+\CO(R).
}

Finally, we use the equations of motion \eom\ to simplify the
formula \pidereval\ for the derivatives of $\Pi_i$, since we will
only be interested in the matrix of coupling constants on-shell.
\eqn\pitwoderred{\eqalign{
2\pi i{\partial \Pi_2 \over \partial f_0}&=-{N_1\over
N\,\sqrt{(r_4-r_2)(r_3-r_1)}}\,F({\pi\over2}|R)\cr 2\pi i{\partial
\Pi_2 \over \partial f_1}&=-{N_1\over
N\,\sqrt{(r_4-r_2)(r_3-r_1)}}
\left(r_1\,F({\pi\over2}|R)+(r_2-r_1)\,\Pi(n;{\pi\over2}|R)\right).\cr
}}
As we approach the singularity, we can use \eomsimp\ to replace
$\theta\rightarrow {\pi\over2}{N_2\over N}$. Moreover
$F({\pi\over2}|R)$ and $\Pi(n;{\pi\over2}|R)\rightarrow
{\pi\over2}$ as $R\rightarrow 0$. Using these facts and
\pitwoderred\ and \kapparew\ in \chain, we obtain at last our
formula for ${\partial\Pi_2\over\partial S_1}$:
\eqn\chainrew{\eqalign{
    {\partial \Pi_2 \over \partial S_1}
    &=
    {\pi i N_1 N_2\over N^2}{1\over \log\left({16\over R}\right)}+\CO(R).\cr
}}

\appendix{B}{The ${\bf Z}_N$ symmetry of the $\CN=1$ AD point}

In section 4.1, 
we gave a heuristic derivation of the ${\bf Z}_N$ symmetry of the
$\CN=1$, $n=n_{min}$ AD points by using the microscopic $U(1)_R$
symmetry. In this appendix, we will derive the ${\bf Z}_N$
symmetry directly from the matrix model calculation of the
glueball superpotential. As before, we must consider separately
the cases of $N$ even and $N$ odd. We first consider the case of
$N$ odd.

To begin, notice that for $N$ odd, the branch points of $y_m$ are
located at $x_0=0$ and at the $N$th roots of unity $x_j=e^{2\pi
i\, (j-1)/ N}$ (we temporarily set $4\Lambda^N=1$ for
convenience). Let us define the $A$-cycles to be $A_1=(0,1)$ and
$A_i=(x_{2i-2},x_{2i-1})$. Then the branch points and the
$A$-cycles possess a natural ${\bf Z}_N$ symmetry, which we will
take to be generated by $x\rightarrow e^{4\pi i/N}\,x$. The matrix
model curve \mmnodd\ transforms as $y_m^2\rightarrow e^{4\pi
i/N}\,y_m^2$ under this transformation if we take the coefficients
of $g_{n-1}(x)$ to transform with charges
\eqn\chargesNodd{
g_j \rightarrow e^{4\pi i\,(1-j)/N}g_j.
}
Using the ${\bf Z}_N$ symmetry and various contour integrals, one
can show that the periods of the one-form $T(x)$ are $N_1=1$ and
$N_i=2$, $b_i=1$ for $i>1$. Thus the effective superpotential
takes the following form for $N$ odd:
\eqn\weffNodd{
    W_{eff}=\Pi_1+\sum_{i=2}^{n}(2\Pi_i+S_i).
}
We have dropped the usual term proportional to the bare gauge
coupling, as this merely serves to renormalize
$\Lambda_0\rightarrow\Lambda$.

In order to realize the ${\bf Z}_N$ symmetry explicitly in the
effective superpotential, it will be useful to imagine deforming
each cut $A_i$ so that it nearly touches the origin (recall that
$A_i$ runs between adjacent $N$th roots of unity). Then, at least
for the purposes of computing the periods $\Pi_i$, we can consider
the $x$ plane as consisting of $N$ ``cuts'' that run from $x=0$ to
the $N$ roots of unity. Define the contours $\tilde{B}_j$,
$j=1,\dots,N$ which connect $\Lambda_0$ on the two sheets and run
through the $j$th ``cut'', and define the periods $\tilde{\Pi}_j$
of $R(x)$ based on these contours:
\eqn\tildepi{
    \tilde{\Pi}_j=-{1\over2}\oint_{\tilde{B}_j}\sqrt{x(x^N-1)+g_{n-1}(x)}\,dx.
}
By various contour deformations, one can relate the original
periods $\Pi_i$ to the $\tilde{\Pi}_j$, and one can show that the
effective superpotential \weffNodd\ simplifies to the following
suggestive form:
\eqn\weffsymmNodd{
W_{eff}= \sum_{j=1}^{N}\tilde{\Pi}_j.
}
From the definition \tildepi, it is clear that the ${\bf Z}_N$
transformation acts on the periods $\tilde{\Pi}_j$ as
\eqn\pitransfNodd{
    \tilde{\Pi}_j\left(g\right) \rightarrow e^{6\pi i/N}\times\tilde{\Pi}_{j+2}\left(g\right).
}
It follows that $W_{eff}$ transforms with definite charge under
the ${\bf Z}_N$ symmetry:
\eqn\wefftransfNodd{
    W_{eff}\rightarrow e^{6\pi i/N}W_{eff}.
}
This confirms the heuristic derivation of the ${\bf Z}_N$ symmetry
given in section 4.1 
for $N$ odd.

The calculation is very similar for $N$ even, so we omit most of
the intermediate steps. The $N$ branch points of the matrix model
curve are located at the $N$th roots of unity $x_j=e^{2\pi i\,
(j-1)/ N}$. We define the $A$-cycles to be
$A_i=(x_{2i-1},x_{2i})$. Then ${\bf Z}_N$ symmetry again acts as
$x\rightarrow e^{4\pi i/N}\,x$, but this time the matrix model
curve \mmneven\  is invariant, as long as we transform the
coefficients of $g_{n-1}(x)$ with charges
\eqn\chargesNeven{
g_j \rightarrow e^{-4\pi i\,j/N}g_j.
}
The periods of $T(x)$ are $N_i=2$ and $b_i=0$, so the effective
superpotential takes the form
\eqn\weffNeven{
    W_{eff}=2\sum_{i=1}^{n}\Pi_i
}
where again we omit the bare-coupling term. Once again, we
consider deforming the cuts so that they nearly touch the origin,
and we define the periods $\tilde{\Pi}_i$, $i=1,\dots,N$. Similar
arguments as for $N$ odd lead to the following expression for
$W_{eff}$:
\eqn\weffsymmNeven{
    W_{eff} =\sum_{j=1}^{N}\tilde{\Pi}_{j}-S.
}
Now the ${\bf Z}_N$ transformation acts on $\tilde{\Pi}_j$ as
\eqn\pitransfNeven{
    \tilde{\Pi}_j\left(g\right) \rightarrow e^{4\pi i/N}\times\tilde{\Pi}_{j+2}\left(g\right).
}
Moreover, since $S=-{1\over4}g_{n-1}$, $S$ also transforms as
$S\rightarrow e^{4\pi i/N}S$ under the action of ${\bf Z}_N$.
Therefore $W_{eff}$ again transforms with definite charge under
the ${\bf Z}_N$ symmetry:
\eqn\wefftransfNeven{
    W_{eff}\rightarrow e^{4\pi i/N}W_{eff}
}
as expected from the arguments of section 4.1. 

\listrefs
\end